\documentclass[%
 aip,twocolumn, 
 amsmath,amssymb,
 reprint,%
]{revtex4-1}

\usepackage[pdftex]{graphicx}

\usepackage{amsmath}
\usepackage[latin1]{inputenc}
\usepackage{txfonts}
\usepackage{hyperref}
\usepackage[pdftex]{graphicx}

\begin{document}

\title{Optimal global synchronization of partially forced Kuramoto oscillators\\$\ $ }

\author{Joyce S. Climaco}
\email{joyce.sclimaco@gmail.com}
\author{Alberto Saa}
\email{asaa@ime.unicamp.br}
\affiliation{
Department of Applied Mathematics, University of Campinas, \\
   13083-859 Campinas, SP, Brazil.}

\date{\today}

\begin{abstract}
We consider the problem of global synchronization in a large random network of Kuramoto oscillators where some of them are subject to an external periodically driven force. We explore a recently proposed dimensional reduction approach and introduce an effective two-dimensional description for the problem. From the dimensionally reduced model, we obtain   analytical predictions for some critical parameters necessary for the onset of a globally synchronized state in the system. Moreover, the low dimensional model also allows us to introduce an optimization scheme for the problem. Our main conclusion, which has been corroborated by exhaustive   numerical simulations, is that for a given large random network of Kuramoto oscillators, with random natural frequencies $\omega_i$, such that a fraction of them is subject to an external periodic force with frequency $\Omega$, the best global synchronization properties correspond to the case where the fraction of the forced oscillators is chosen to be those ones such that $|\omega_i-\Omega|$ is maximal. 
Our results might shed some light on the structure and evolution   of
natural systems for which the presence or the absence of global synchronization are desired
properties. Some properties of the optimal forced networks  and its relation to   recent results in
the literature 
are also discussed. 
\end{abstract}
 
 
\maketitle

\begin{quotation}
We consider here the dynamics of a large number of interacting Kuramoto oscillators.
 Each oscillator
has its own natural frequency $\omega_i$, which is assumed to be a random variable, and the interactions
among them are associated with the edges of a random network. Despite of being essentially a  
random system, there are plenty of robust results obtained from this kind of model which have proven
to be relevant in 
   many different areas. This is the case, for instance, of synchronization phenomena,  see   \onlinecite{Review1} and \onlinecite{Review2}  for   comprehensive reviews on the subject. Here, we are concerned with one of the main variations of the network of Kuramoto oscillators,  the case
   where some of the oscillators are subject to an external periodically driven force with frequency $\Omega$. By exploring a recently proposed analytical approach, we introduce an optimization scheme for the onset of the so-called global synchronization in the system, a regime where all oscillators rigidly rotate, forming a compact swarm, in the same pace of the external force,  with frequency $\Omega$. We show that the best global synchronization properties correspond to the case where the set of 
     forced oscillators is chosen to be those ones such that the value of $|\omega_i-\Omega|$ is maximal.  Our results may help to  understand the evolution and structure of   natural systems with
     many interacting agents for which    global
 synchronization  plays an important dynamical role. 
\end{quotation}

\section{Introduction}

Synchronization   in complex networks of oscillators is a paradigmatic  dynamical  
problem which
 has received huge attention recently.
Its intrinsically  rich dynamics and vast applicability range have motivated   a myriad of studies
in very different areas, see 
Refs.   \onlinecite{Review1} and \onlinecite{Review2}  for   comprehensive reviews on the subject.  The mostly used model for synchronization studies  is still the well-known 
Kuramoto  oscillator, introduced more than   forty years ago \cite{kuramoto1975} in the context
of chemical oscillations. The list of situations involving interacting individual 
agents which can be well described dynamically 
by a network of Kuramoto oscillators is vast, ranging from Biology to mass communication technologies\cite{Review2}. In many cases, the real underlying interaction network is so large and entangled that any intention to model it will necessary require some kind of random description, typically with certain prescribed statistical properties for the nodes and interactions among them\cite{Newman}.    Roughly, this is namely
the key concept of a complex network for our purposes: a random graph with   certain prescribed statistical properties for its vertices and edges attributes. The connection pattern and the frequencies distribution in a network of Kuramoto oscillators are known to affect its synchronization properties,
and since in many applications
 synchronization
is a dynamical regime expected to be favored (or suppressed), many recent works have been devoted to the
study of the optimization of synchronization in complex networks\cite{pecora1988,barahona2002,m1,m2,m3,donetti2005,dwivedi2015,PintoSaa,optimal,AISync}. Such
optimization schemes are useful not only to enhance the synchronization capability of a given complex
network by adjusting its connection topology, but also to understand the evolution and structure of natural systems with
many interacting agents for which   synchronization
 plays a fundamental role.

One of the main variations of the  classical   
  synchronization problem 
is   the case of   
  a network of Kuramoto oscillators    subject to  
  external periodically driven forces\cite{p1,p2,p3,p4,p5,p6,p7}. The dynamics 
  become much richer in such a case, 
  with distinct notions of synchronization available and the appearance of possible bifurcations among them.  Here, we
 consider another variant of the classical problem: the case where only a fraction of the oscillators is
    subject to  
the external periodical  force. There are plenty of motivations in natural systems and many recent examples of application involving networks of partially forced Kuramoto oscillators. In Biology, for instance, there are several situations
where the response of specific groups of cells are triggered by external stimuli acting on other
cells. Typical examples include the receiving and processing of visual, olfactory, or haptic signals, but the list is certainly much larger. The
dynamical models for these cases consist commonly in a complex network where the cells, which pertinent
dynamical states are described by   Kuramoto oscillators, are located at the vertices, with the edges corresponding to the interaction among them. Recent applications of 
partially forced Kuramoto oscillators include  the study of the response to external stimuli
of the C. elegans neural network\cite{modular, p8} (see   \onlinecite{book1} for a review  on 
complex network 
models for 
  the response of neural networks to external stimuli), 
self-organization and pattern formation
in the growing and developing  of vertebrates\cite{p9},  auditory signals in amphibians\cite{p10}, and several issues on the
circadian rhythm\cite{p11,p12,p13}.  

The problem we are concerned here corresponds to 
 a connected network with $N$ Kuramoto oscillators  
 governed by  following dynamical equations 
\begin{equation}
\frac{d \theta_{i}}{dt} = \omega_i + \mathbf{1}_{C}(i)F\sin\left(\Omega t - \theta_i \right) + \lambda \sum_{j=1}^{N} A_{ij} \sin(\theta_j - \theta_i),
\label{kuramoto}
\end{equation}
where $\theta_i$ stands for the dynamical state (phase) of the Kuramoto oscillator with
natural frequency $\omega_i$, which is assumed to be located at the $i^{\rm th}$-node of the 
underlying 
network.  
The  connections among the oscillators  are represented    by the usual
undirected  adjacency matrix with components $A_{ij}$, 
 and $\lambda$ stands for the  uniform coupling strength among the oscillators. We will consider here
 the so-called attractive case, for which $\lambda >0$. 
The subset of $N_C<N$ forced nodes is denoted by $C$, and $\mathbf{1}_{C}(i)$ is its indicator function, {\em i.e.},
 $\mathbf{1}_{C}(i) = 1$
  if $i\in C$, or zero otherwise. The external driven force is also assumed to have uniform intensity $F$ and frequency $\Omega$ for all nodes   $i\in C$. Without loss of generality, we can assume 
$F\ge 0$. In fact, if $\theta_i(t)$ is a solution of (\ref{kuramoto}) for some
external force $F$, $\theta_i(t) + \pi$ will be  a solution for the case corresponding to $-F$, and
both cases have  the same asymptotic dynamical properties. 
 We will also assume  $\Omega > 0$ in our analysis,
but we will see that the case $\Omega < 0$ can be treated analogously. 

Our analysis  will be   
  restricted  to  the situations where the underlying network is an undirected and unweighted
large 
random network. We assume also that $\omega_i$   is a random variable over the
network, with null average, symmetric 
distribution $g(\omega)$,  and all moments finite. 
  Notice that we could also have written the system (\ref{kuramoto}) in an autonomous manner by introducing an extra node $ (N+1)$ with natural frequency $\Omega$ and connected in a directional way to the forced nodes $i\in C$. For a recent discussion on the dynamics of
   directed and weighted networks of Kuramoto oscillators, see
  \onlinecite{p14}.  However, since our proposed approach is indeed more appropriate to the case of undirected and unweighted
large 
random networks,   we will deal with the original non-autonomous formulation of the problem.  
There are several dynamical behaviors for the system (\ref{kuramoto}) which would deserve
to be named synchronization. Since the external driven force has its own frequency $\Omega$, we will
focus here on the network dynamical states for which the oscillators rigidly rotate
in the same pace of the external force, 
 with frequency $\Omega$, and in a phase-locked way. This situation is also called in the literature as the global synchronization for (\ref{kuramoto}). Of course, it is quite natural to expect  the existence of a certain threshold  $F_{\rm c} > 0$  for
the  external driven force such that, for $ F <F_{\rm c}$, the
system (\ref{kuramoto})  would be rather insensitive to the external excitation. In fact, this is a well-known property of the system (\ref{kuramoto}) for the case of a fully connected network topology, see  \onlinecite{p3} 
for further details. 

We will employ  the dimensional reduction approach recently proposed
by Gottwald \cite{gottwald2015}, which in turn  is   based in the introduction of some collective coordinates for the Kuramoto oscillators in the same spirit of the Ott-Antonsen ansatz \cite{p2,OA1,OA2}, to investigate
the global synchronization properties of (\ref{kuramoto}) for the case of large random networks. 
The Gottwald approach was recently extended for the case of Stuart-Landau complex oscillators in \onlinecite{AISync} to incorporate also amplitude effects in the dynamics. From the dimensionally reduced system, 
we will derive some conditions on the parameter space of the problem
 which will allow (or prevent) the onset of a globally
synchronized state for the system. Moreover, in the same line of the rewiring algorithm discussed in
\onlinecite{optimal,AISync}, we will explore the dimensional reduction approach to propose an optimization scheme
for global synchronization by selecting judiciously the forced nodes subset $C$ for 
a given random network. The scheme has been
exhaustively tested with numerical simulations and it has always resulted in an enhancement in
the global synchronization properties of the system. Our main conclusion is that the optimal subset $C$, in the sense that will be properly define in Section \ref{section3}, should consist  in the nodes $i$ such that $|\Omega -\omega_i|$ is maximal, a result which is indeed in agreement  with some recent related works \cite{optimal,AISync,SciRep}, and that 
 might shed some light on the   evolution and structure of   natural systems for which global
 synchronization  is a desired property. Despite our analysis is explicitly 
  done for the $\Omega>0$ case
 with symmetric distributions $g(\omega)$, we will show that the optimization criterion of maximizing $|\Omega -\omega_i|$
 for choosing the subset $C$ of forced nodes is   valid for more general cases as well. 

The present paper is organized as follows. In the next section, the Gottwald  dimensional reduction approach\cite{gottwald2015}  
is adapted for the global synchronization problem   of partially forced Kuramoto oscillators
governed by (\ref{kuramoto}). A mean field analysis is performed, and  we derive some conditions on the parameter  space
of the
problem 
for the onset of a
globally synchronized state. 
In Section \ref{section3}, we introduce our optimization scheme   and present the results of our numerical
simulations for  large random networks. 
The last section is devoted to some concluding remarks on the implications of
our results and on the role played by possible network symmetries
on the global synchronization problem of forced Kuramoto oscillators,  in the context of the
recently introduced  
asymmetry-induced synchronization (AISync) scenario\cite{
NishMotter, ZhangNishMotter, ZhangMotter}.    

\section{The dimensional reduction approach}

 We will describe the global state of the system (\ref{kuramoto}) by using   the standard
  order parameters $r$ and
 $\psi$  
 defined as
\begin{equation}
r(t)e^{i\psi(t)}  =   \frac{1}{N}\sum_{j=1}^{N}e^{i\theta_j(t)} ,
\label{order_parameter}
\end{equation}
whose respective behaviors are well known: 
 $r \approx N^{-1/2}$ for incoherent motion, whereas $r \approx 1$ for a fully synchronized 
state.
Our global synchronization regime corresponds to a 
  phase locked configuration    with a
rigid rotation $\Omega$, and hence to the case $r \approx 1$  and  $\dot\psi \approx \Omega$.   
We will look for globally synchronized states of (\ref{kuramoto}) by exploring
a two dimensional model based on   the Gottwald dimensional reduction approach introduced in
 \onlinecite{gottwald2015}. Gottwald ansatz is based on the empirical observation that the
 set of synchronized Kuramoto oscillators in a complex network typically have their phases 
 obeying $\theta_i \propto \omega_i$, up to a rigid rotation with frequency $\langle\omega \rangle$. By considering the
 collective ansatz $\theta_i(t) = \mu(t)\omega_i + \langle\omega\rangle$, we see that
 synchronization corresponds to the existence of a fixed point for $\mu(t)$ for large $t$, see
 \onlinecite{optimal} for further details. 
 In the present case, the Gottwald approach 
 can be adapted as the 
collective ansatz
\begin{equation}
\label{Gottwald}
\theta_i(t) = \alpha(t) + \frac{\omega_i}{\Omega}\beta(t) + \Omega t 
\end{equation}
for the entire network, 
which has been indeed extensively tested in our numerical simulations, as we will discuss in the next
section. Notice that, 
by construction, both reduced dynamical variables $\alpha$ and $\beta$ are dimensionless. 
Using such ansatz and
recalling that $\omega_i$ is assumed to be a random variable with null average and symmetric distribution $g(\omega)$,
 we will have for large random networks 
  \begin{equation}
\label{rfixed}
r(\beta)  = \left\langle \cos \frac{\beta \omega}{\Omega} \right\rangle = \int d\omega \, g(\omega) \cos  
\frac{\beta \omega}{\Omega}  
\end{equation}
 and
\begin{equation}
\label{psifixed}
\psi = \alpha + \Omega t,
\end{equation}
from where the dynamical interpretation of $\alpha$ and $\beta$ will become rather clear. 
We denote the simple average of a given variable $h_k$ over the entire network as
\begin{equation}
\label{averagdef}
\langle h \rangle = \frac{1}{N}\sum_{k=1}^Nh_k,
\end{equation}
and the continuous (mean field) approximation consisting basically in  the substitution of the
sum with the integral with the pertinent probability distribution, as we have done in (\ref{rfixed}).
Notice that synchronized states with $r\approx 1$ will demand  $\beta\approx 0$. In fact,
the expression (\ref{rfixed}) can be expanded as
\begin{equation}
r\approx 1 - \frac{\langle \omega^2 \rangle}{2\Omega^2} \beta^2 + \frac{\langle \omega^4 \rangle}{4!\Omega^4} \beta^4 + \cdots,  
\end{equation} 
for any distribution $g(\omega)$ with finite moments. 
It is clear that   asymptotic  globally  synchronized states will correspond to 
solutions such that $\beta \to 0$ and $\alpha \to $ constant for $t\to\infty$. Notice that
the ansatz (\ref{Gottwald}) can also accommodate the description of a  synchronized regime rigidly rotating with frequency $\Omega'\ne\Omega$. This case would correspond  to a solution with $\beta \to 0$ and
$\alpha \approx (\Omega' - \Omega)t$ for large $t$.

\begin{figure}[t]
\includegraphics[scale=0.5]{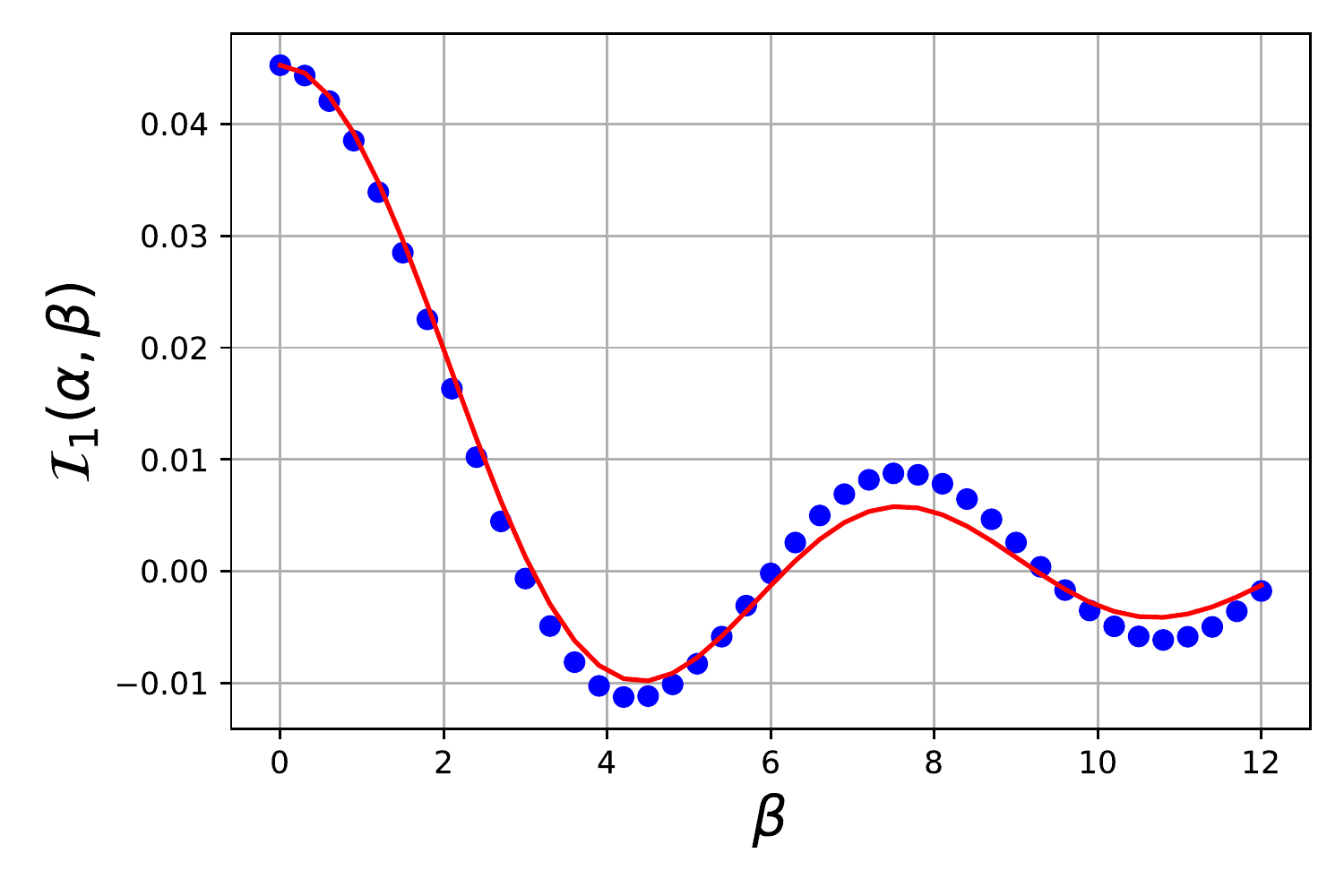} 
\caption{The aspect of the function ${\cal I}_1(\alpha,\beta)$ defined by (\ref{I_1}) for 
an Erd\"os-R\'enyi network with 1500 nodes,
 average degree $\langle d\rangle \approx 7.7$ and
 with only 96 nodes in the set $C$, with a uniform frequency distribution over
 $(-1,1)$.  
 The read line is the mean-field prediction (\ref{aprox1}),  and the blue circles   correspond to the respective numerical values calculated from (\ref{I_1}). The depicted case corresponds to 
 $\alpha = \frac{\pi}{4}$, but the overall accuracy is rather insensitive to the specific value
 of $\alpha$. On the other hand, the value of $N_C$ does play an important role. The larger $N_C$, the most accurate is the  approximation  (\ref{aprox1}). The situation for ${\cal I}_2(\alpha,\beta)$ defined by (\ref{I_2}) is analogous. For the case of ${\cal I}_3(\beta)$ defined by (\ref{I_3}), see \onlinecite{AISync}.
}
 \label{Fig0}
\end{figure}

In order to obtain the two-dimensional reduced system in the new variable $\alpha$ and $\beta$, let us 
substitute   (\ref{Gottwald})  in (\ref{kuramoto}), which will result  in
\begin{eqnarray}
\label{eq3}
\dot\alpha + \frac{\omega_i}{\Omega}\dot\beta &=& \omega_i-\Omega - \mathbf{1}_{C}(i)F \sin\left( \alpha + \frac{\omega_i}{\Omega}\beta  \right) \\
&& \quad\quad\quad  + \lambda \sum_{j=1}^{N} A_{ij} \sin\frac{\beta}{\Omega}(\omega_j - \omega_i). \nonumber 
\end{eqnarray}
Averaging both sides according to (\ref{averagdef}) leads to
\begin{equation}
\label{eqa}
\dot\alpha =   -\Omega -  F  {\cal I}_1(\alpha,\beta),
\end{equation}
where the dimensionless function ${\cal I}_1(\alpha,\beta)$ is given by
\begin{equation}
\label{I_1}
{\cal I}_1(\alpha,\beta) = \frac{1}{N}\sum_{i=1}^N\mathbf{1}_{C}(i)\sin\left( \alpha + \frac{\omega_i}{\Omega}\beta  \right).
\end{equation}
On the other hand, multiplying Eq. (\ref{eq3}) by $\frac{\omega_i}{\Omega}$ and averaging both sides again, we get
\begin{equation}
\label{eqb}
  \dot\beta =  \Omega - \frac{F \Omega^2}{ \langle \omega^2 \rangle } {\cal I}_2(\alpha,\beta) 
  + 
  \frac{\lambda\Omega^2}{  \langle \omega^2 \rangle} {\cal I}_3(\beta) ,
\end{equation}
where 
\begin{equation}
\label{I_2}
{\cal I}_2(\alpha,\beta)  =
\frac{1}{N}\sum_{i=1}^N\mathbf{1}_{C}(i)\frac{\omega_i}{\Omega}\sin\left( \alpha + \frac{\omega_i}{\Omega}\beta  \right)       
\end{equation}
and
\begin{equation}
\label{I_3}
{\cal I}_3(\beta) =
\frac{1}{N}\sum_{i=1}^N\sum_{j=1}^{N} A_{ij}\frac{\omega_i}{\Omega} \sin\frac{\beta}{\Omega}(\omega_j - \omega_i) 
\end{equation}
are also dimensionless functions. 
The continuous (mean field)  approximation in the present case corresponds to 
\begin{eqnarray}
\label{aprox1}
{\cal I}_1(\alpha,\beta)  &=& f r(\beta) \sin\alpha ,  \\
\label{aprox2}
{\cal I}_2(\alpha,\beta)  &=& -f   r'(\beta)  \cos\alpha ,  \\
\label{aprox3}
{\cal I}_3(\beta)  &=&\frac{\left\langle d\right\rangle}{2} \frac{d}{d\beta} r^2(\beta),
\end{eqnarray}
where $f= N_C/N$ is the fraction of forced nodes, $\left\langle d \right\rangle$ stands for the average degree of the network and the prime denotes derivative. 
In the derivation of such approximations, we employ the crucial hypothesis   that the $N_C$ forced nodes were also randomly
chosen and, in particular, that their natural frequencies has the same distribution 
$g(\omega)$. For further details, see the Appendix \ref{appendix}. 
Equations (\ref{eqa}) and (\ref{eqb})   in this case read simply
\begin{eqnarray}
\label{eqa1}
\dot \alpha &=& -\Omega - {fF}r(\beta)\sin\alpha, \\
\label{eqa2}
\dot \beta &=& \Omega + r'(\beta)\frac{\Omega^2}{\left\langle \omega^2\right\rangle}\left(  
 {\lambda\left\langle d \right\rangle }  r (\beta) +  fF  \cos\alpha   \right).
\end{eqnarray} 
It is important to stress that the quantities (\ref{aprox1})-(\ref{aprox3}), which give origin
to the dynamical two-dimensional system (\ref{eqa1})-(\ref{eqa2}), are completely  determined from
the network and oscillators data. For instance, for   normal and   homogeneous distributions
$g(\omega)$,  the most commonly used ones in the literature,   we have    the following continuous approximations for $r$ from (\ref{rfixed})
\begin{equation}
\label{rn}
r_{\rm n}  =
\exp\left( -\frac{ \langle\omega^2  \rangle  \beta^2 }{2\Omega^2}   \right)
\end{equation}
and
\begin{equation}
\label{ru}
r_{\rm u} = \frac{\sin \sqrt{3\left\langle\omega^2 \right\rangle } \frac{\beta}{\Omega}  }{\sqrt{3\left\langle\omega^2 \right\rangle}  \frac{\beta}{\Omega} },
\end{equation}
respectively, 
from where all the quantities (\ref{aprox1})-(\ref{aprox3}) can be derived. The mean-field
approximations are   accurate for large random networks, see Fig. \ref{Fig0}
for a typical example. 

For $r(\beta)=1$, Eq. (\ref{eqa1}) reduces to the well-known Adler equation, exactly  as in the
full-connected topology case, see \onlinecite{p3} and \onlinecite{p8} for further details. Since 
$  0< | r(\beta)|\le 1$, equation (\ref{eqa1}) 
will never admit fixed points if 
\begin{equation}
\left|\frac{fF}{\Omega}\right| < 1.
\end{equation}
We have just established the critical value of the external   force $F_{\rm c}$ for the  
onset of the global synchronization, which will always require
\begin{equation}
\label{F_c}
F > F_{\rm c} = \frac{\Omega}{f}.
\end{equation} 
Of course, our main interest here is the stable fixed points $(\alpha_*,\beta_*)$ of the system
(\ref{eqa1}) - (\ref{eqa2}), which Jacobian matrix for $\beta\approx 0$ reads 
\begin{equation}
\frac{\partial (\dot\alpha,\dot\beta)}{\partial (\alpha,\beta)} =-\left( 
\begin{array}{cc}
 fF \cos\alpha & 0 \\
0 &   
 {\lambda\left\langle d \right\rangle }  +  fF  \cos\alpha   
\end{array}
\right),
\end{equation} 
from where  we have that any stable fixed point corresponding to a synchronized state
should have $fF\cos\alpha_*>0$. One can explore the right-handed side of equation
(\ref{eqa2}) to derive some critical value for $\lambda$ in order to assure a globally synchronized   
state in the same way we did for $F$ in (\ref{F_c}), but it turns out that the situation is very similar
to the case without the external force ($F=0$), which was treated in detail in some previous
works \cite{optimal,AISync}. In particular,  one should not expect any stable fixed point for (\ref{eqa1}) - (\ref{eqa2}) if
\begin{equation}
\label{l_c}
   \frac{\lambda\left\langle d \right\rangle}{\Omega } + \frac{fF}{\Omega} <   {\zeta}, 
\end{equation}
where $\zeta$ is the  constant
\begin{equation}
\zeta = -\frac{ {\langle \omega^2 \rangle}}{\Omega^2\min r'(\beta)} > 0.
\end{equation}
One can evaluate the constant $\zeta$ for   the normal and the homogeneous 
distributions $g(\omega)$. Taking into account (\ref{rn}) and (\ref{ru}),
 we  have $\zeta =  {\sqrt{e}} \approx 1.649$ for the uniform distribution, while for the uniformly distributed case we can numerically determine that $\zeta \approx 2.293$.  However, in contrast with the critical force prediction (\ref{F_c}), which has proved to be indeed quite accurate   in our numerical simulations, the condition (\ref{l_c}) for a possible $\lambda_{\rm c}$ seems to be a rather conservative one. Typically, one will need  larger values for $\lambda$ in order to have robust synchronized states. This situation is completely analogous to the $F=0$ case discussed previously in \onlinecite{optimal,AISync}.

\section{The optimization scheme }
\label{section3}

Our main purpose here is to present a prescription to select judiciously the subset $C$  of forced nodes in order to assure a better synchronization capability for the network. 
We will consider that a network has a better capability for a globally synchronized state if one can attain
states with $r\approx 1$ and $\dot\psi\approx \Omega$ with smaller values for the external force intensity $F$. Since the globally synchronized states require $\beta\approx 0$,   let us linearize 
(\ref{eqa}) and (\ref{eqb}) 
around $\beta = 0$ and abandon the hypothesis that $C$ is a random subset of nodes of the network. We 
will have in this case
\begin{eqnarray}
\label{eqalin}
\dot\alpha &=&   -\Omega -   fF\left( \sin\alpha + \frac{\left\langle \omega \right\rangle_C}{\Omega} \beta \cos\alpha \right), \\
\label{eqblin}
\dot\beta &=&  \Omega \left( 1 -
\frac{  \left\langle \omega \right\rangle_C }{\langle \omega^2 \rangle}fF\sin\alpha \right)  \\
& &\quad\quad \quad
- 
\left( fF\frac{ \langle \omega^2  \rangle_C}{ \left\langle \omega^2\right\rangle  }
 \cos\alpha +    {\lambda } \cal L
 \right) \beta,  \nonumber 
\end{eqnarray}
where $\langle \  \   \rangle_C$ denotes the   average in the subset $C$ and 
\begin{equation}
\label{cal_L}
\mathcal{L} =\frac{ \sum_{e(i,j)}(\omega_i - \omega_j)^2}{\sum_{k=1}^{N}\omega_k^2},
\end{equation}
where the sum in the numerator is performed over the 
 edges $e(i,j)$ of the network. The quadratic quantity $\mathcal{L}$ is known to play
 a crucial role in the usual ($F=0$) synchronization problem \cite{optimal,AISync}.
 In particular, we have that the larger the value of 
 $\mathcal{L}$, the better the synchronization properties of the underlying network, see
 \onlinecite{AISync} for further details.  
 The increasing of $\cal L$  by means of some rewiring operations in the network is the central
 point of the optimization algorithm introduced in \onlinecite{optimal}, which consists basically
 in changing the edges $e(i,j)$ of the network in order to connect oscillators such that $|\omega_i - \omega_j|$ is maximal.

 The fixed points of (\ref{eqalin}) 
 and (\ref{eqblin}) are such that
\begin{equation}
\label{linear}
\frac{\beta_*}{\Omega} = \frac{1 + \frac{ \langle \omega   \rangle_C }{ \langle \omega^2  \rangle }\Omega }
{fF\left( \frac{ \langle \omega ^2  \rangle_C }{ \langle \omega^2  \rangle } -
\frac{ \langle \omega   \rangle_C }{ \Omega }
\right)\cos\alpha_*+\lambda\mathcal{L}},
\end{equation}
from where we can   see that one effectively  attains better values for $r$ (close to 1, requiring smaller
$\beta_*$) if $\langle \omega ^2  \rangle_C$ is maximal and $\langle \omega   \rangle_C$ is minimal. 
This is equivalent to select
the subset $C$ of forced nodes as those ones with the minimal values of their frequencies  $\omega_i$
 in the network. Since we are assuming $\Omega>0$, one can say that
  the optimal subset $C$ is formed by the oscillator such that
 $|\omega_i-\Omega|$ is maximal. As we will see, this criterion is also valid for the  $\Omega < 0$ case
 and even for more general distributions $g(\omega)$.  Notice   that, from (\ref{linear}), we have that $\lambda\cal L$ also
 plays an important role here. As in the $F=0$ case,   the larger the value of 
 $\lambda \mathcal{L}$, the larger the value of $r$ (smaller $\beta_*$). On the other hand, the
 system is rather insensitive to the precise value of $\alpha_*$. Our predictions obtained from the dimensionally reduced
 model, including the optimization criterion, were exhaustively tested in numerical simulations,  whose
main details will be presented in the following subsection.

\subsection{Numerical results}
\label{subsection}
For our simulations, we have numerically solved the equations (\ref{kuramoto}) for large random networks and construct several synchronization diagrams. We have made  extensive use of the 3.7 Python
packages  NetworkX\cite{NetworkX}, which allows us to build many
types of random networks with some prescribed topological and statistical properties, and 
 SciPy\cite{SciPy},   particularly its {\tt integrate.solve\_ivp} function. For our purposes here, it is more convenient to introduce the dimensionless 
 evolution parameter $\tau = \Omega t$ in (\ref{kuramoto}), leading to the following  system of 
 ordinary differential equations 
 \begin{equation}
 { \theta}_{i}'= \frac{\omega_i}{\Omega} + \mathbf{1}_{C}(i)\frac{F}{\Omega}\sin\left(\tau - \theta_i \right) + \frac{\lambda}{\Omega} \sum_{j=1}^{N} A_{ij} \sin(\theta_j - \theta_i),
\label{kuramoto-tau}
 \end{equation}
 where the prime denotes differentiation with respect to $\tau$. As we can see, for a given 
 random network with adjacency matrix $A_{ij}$ and oscillator frequencies $\omega_i$, our
 parameter space is effectively bi-dimensional and spanned by $(F/\Omega, \lambda/\Omega)$. In this case,
 $\Omega^{-1}$ simply plays  the  role of the unit of time for the problem. 
 The equations (\ref{kuramoto-tau}) 
 are the base of all our numerical analysis. We perform the simulations for different network topologies 
 and frequencies distributions $g(\omega)$ with null average, and we have not detected any
 appreciable  dependence
 of our results on the network topology or the employed distribution  $g(\omega)$. All results presented
 here correspond to the case of an  Erdos-Renyi random network with 1500 nodes, 
average degree $\langle d\rangle \approx 7.7$ (the maximal node degree is 18, the minimal is 1), and
with 151 nodes subjected to the external force. We employ a normal distribution 
  $g(\omega)$  with $\sigma_\omega/\Omega = 1$ for the oscillators natural frequencies. None of these parameters has demonstrated to
  have strong influence
  on the results. 
The initial conditions $\theta_i(0)$ for the numerical solutions of (\ref{kuramoto-tau})  were
 randomly chosen,
with uniform distribution, in the circle $[0,2\pi)$.

\begin{figure}[t]
\includegraphics[scale=0.5]{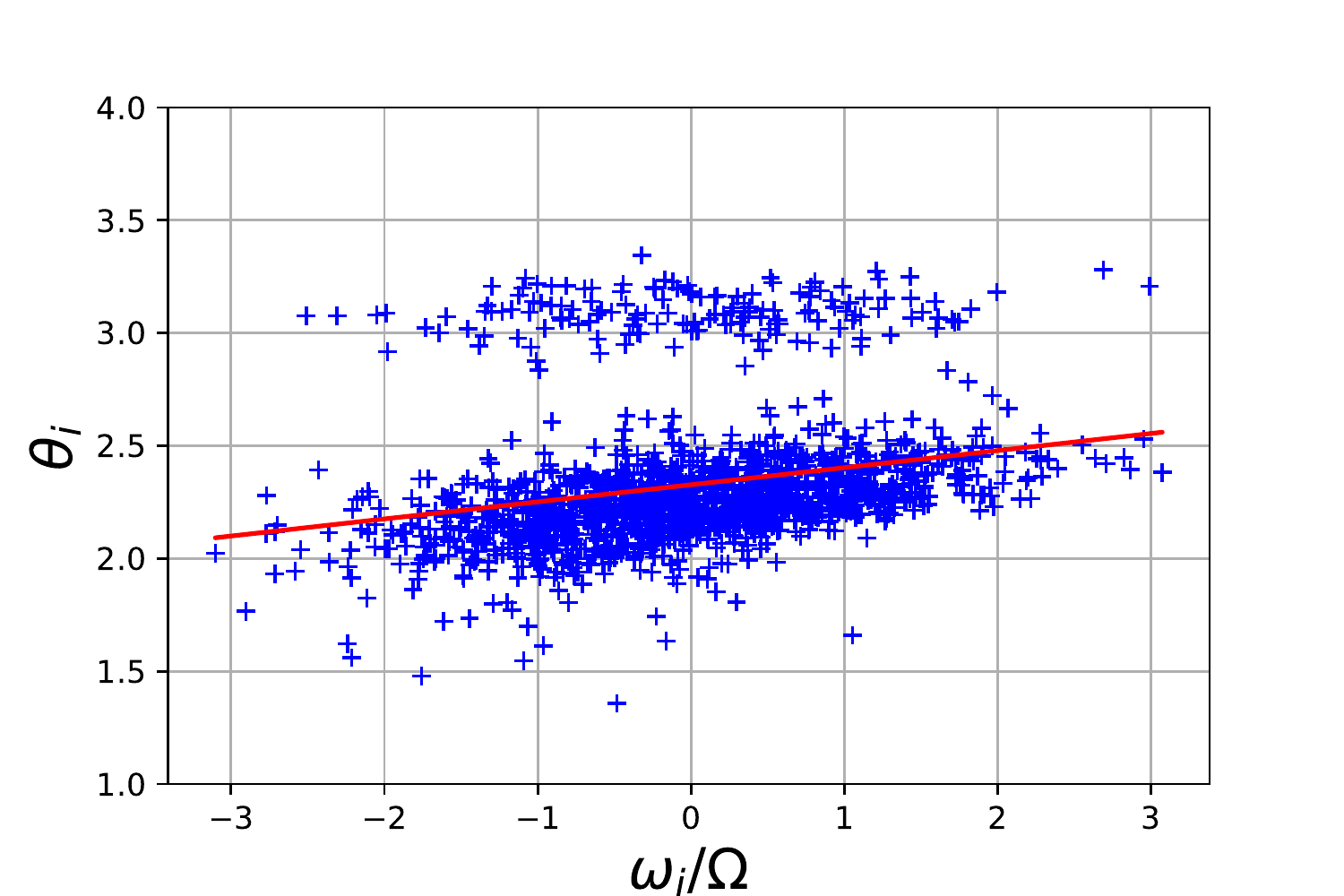} 
\caption{The $(\omega_i,\theta_i(\tau)\mod 2\pi)$ graphics  for a fixed $\tau$
 in a globally
 synchronized regime, for an  Erdos-Renyi random network with 1500 nodes,
with 151 of them subjected to the external periodic force.  
Each point in the graphics corresponds
 to a Kuramoto oscillator in the network. The drawn line is the simple linear regression of the data.
  The linear correlation between
 $\omega_i$ and $\theta_i$, as incorporated in the ansatz (\ref{Gottwald}), is evident. 
  Moreover, we can clearly 
 identify two displaced oscillator populations which similar linear regression slopes.
See the last section for further details on this curious dynamical behavior. }
 \label{Fig1}
\end{figure}

 Our first results, depicted in   Fig. \ref{Fig1},
 correspond to the verification that the ansatz (\ref{Gottwald}) is indeed valid for the study of
 global synchronization in the system (\ref{kuramoto-tau}).
 \begin{figure*}[t]
\includegraphics[scale=0.5]{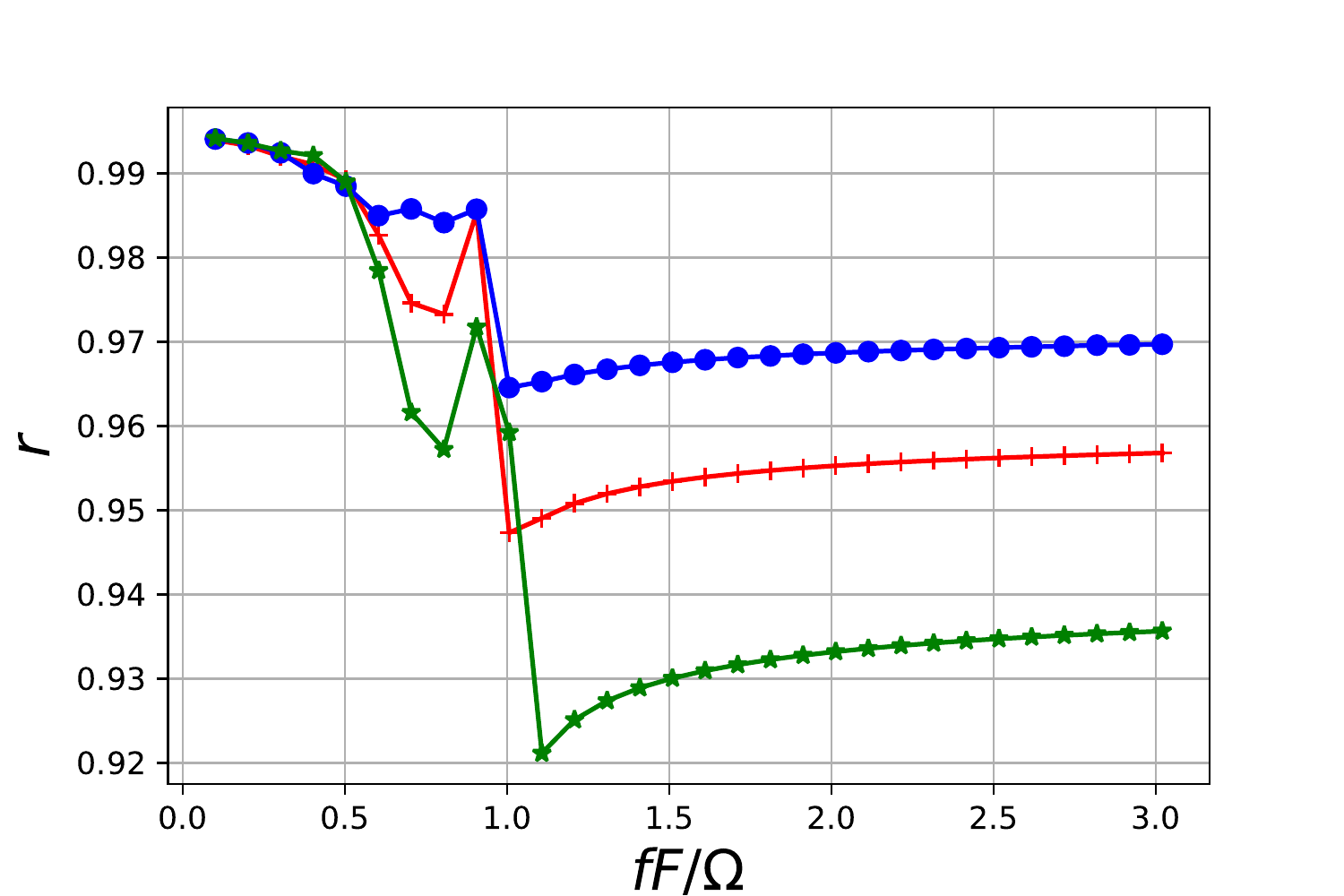}  
\includegraphics[scale=0.5]{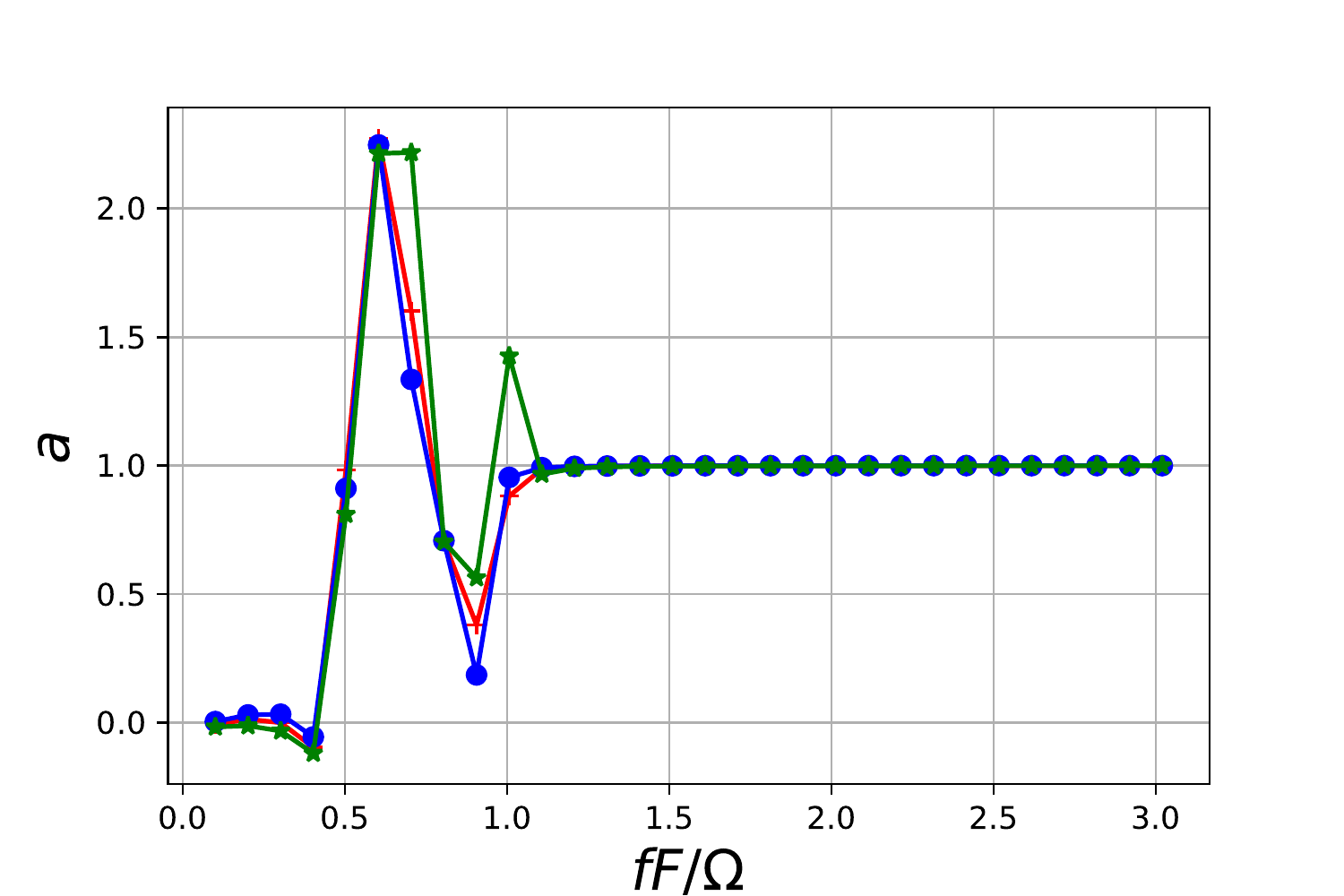}  
\includegraphics[scale=0.5]{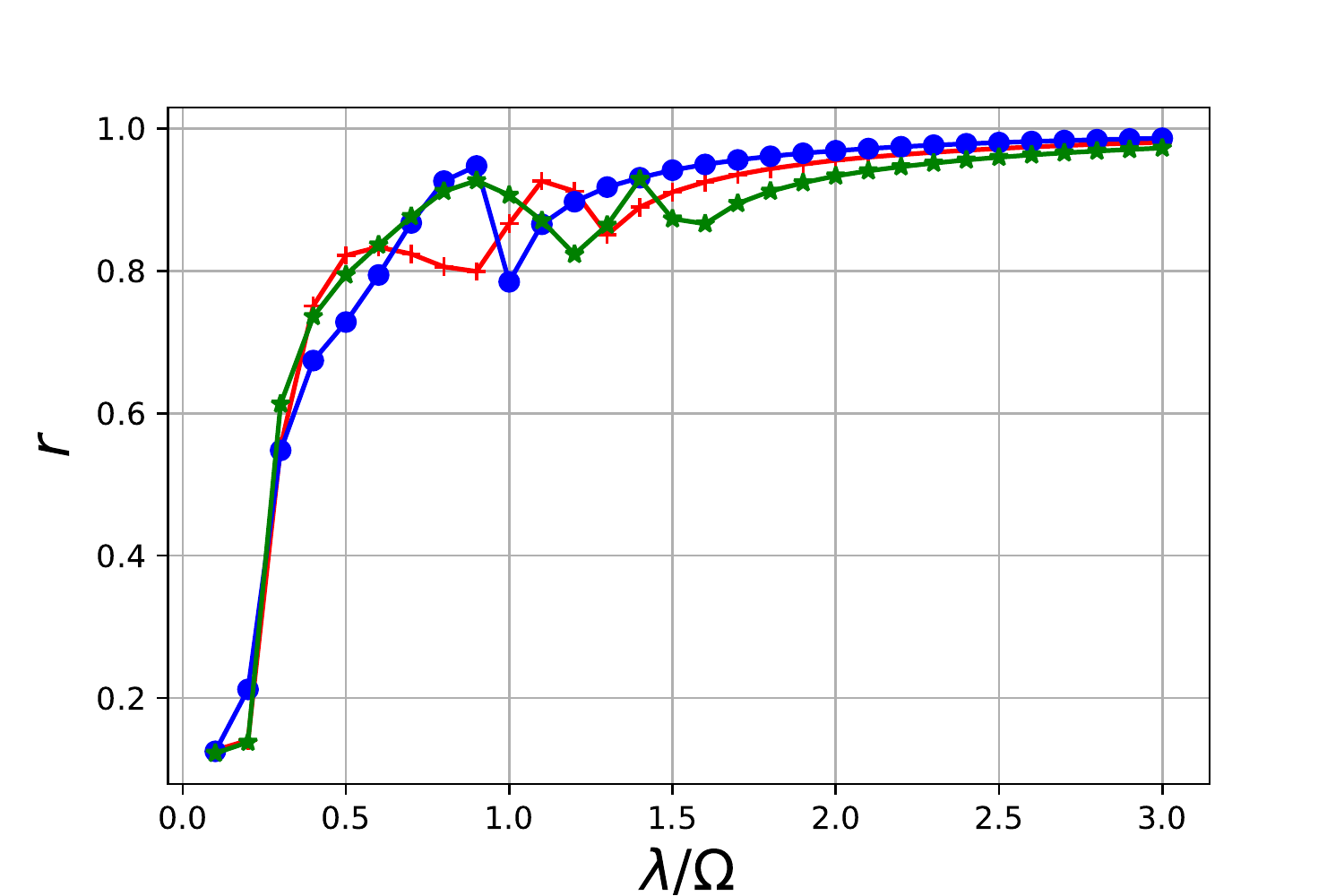}  
\includegraphics[scale=0.5]{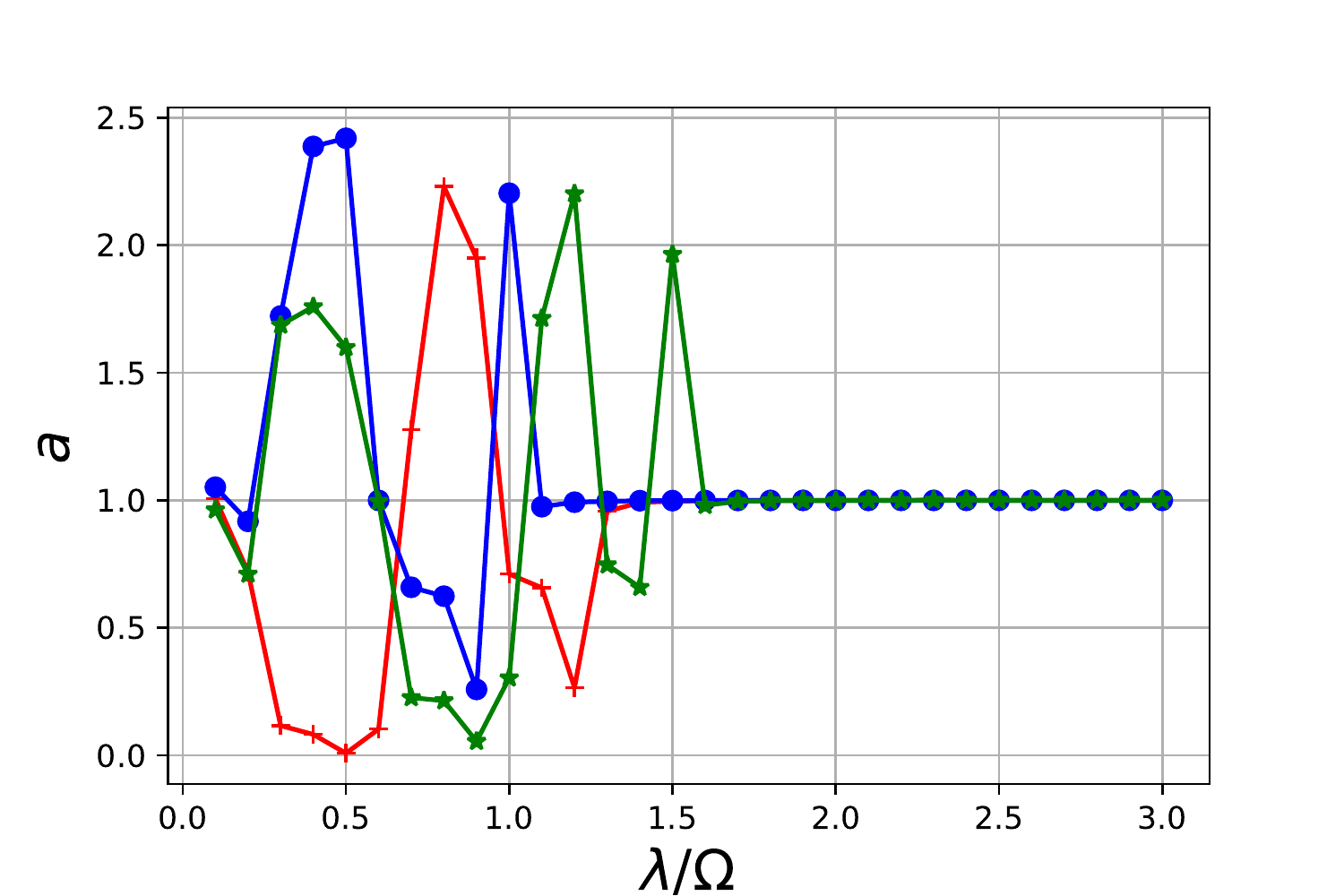}  
\includegraphics[scale=0.5]{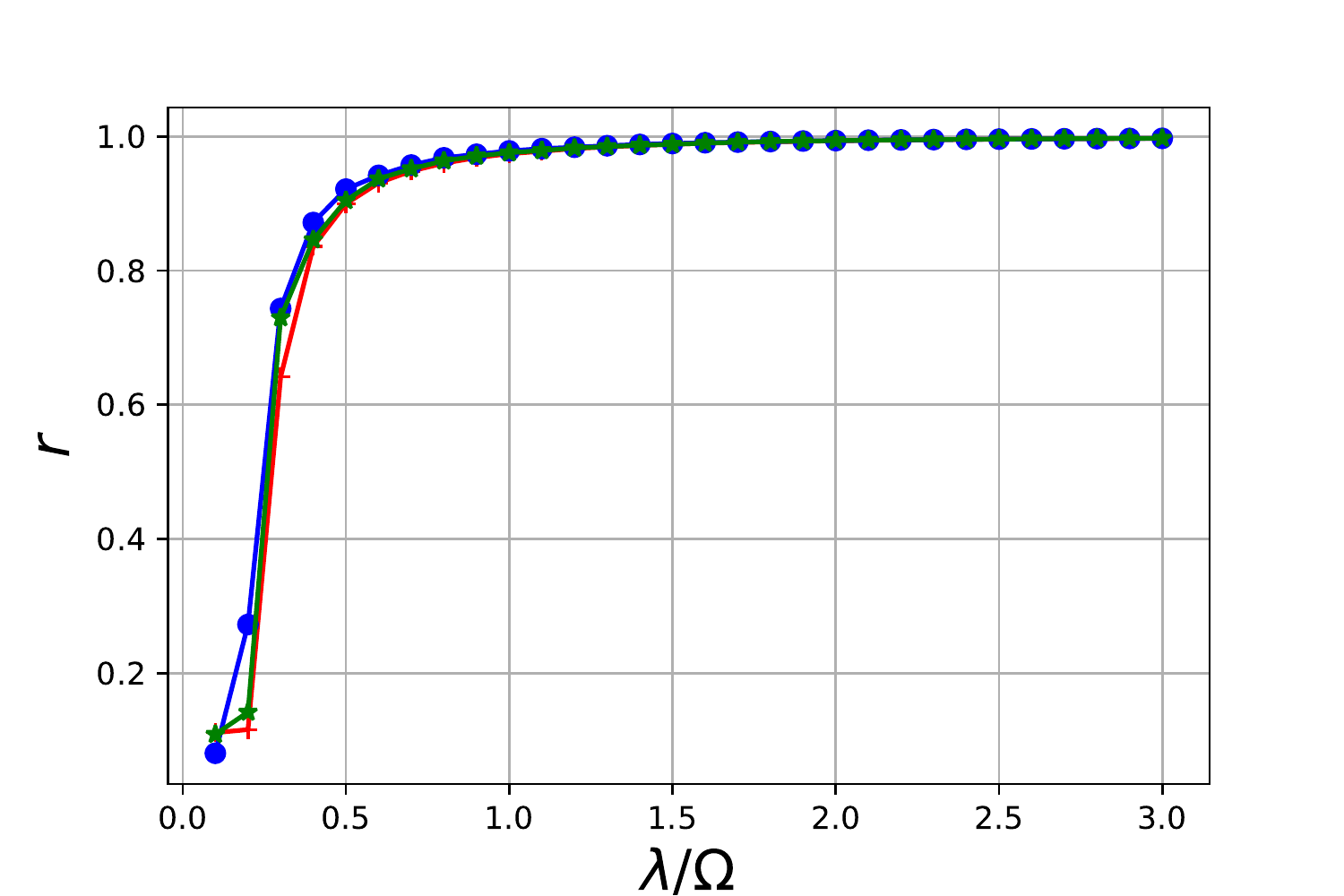}  
\includegraphics[scale=0.5]{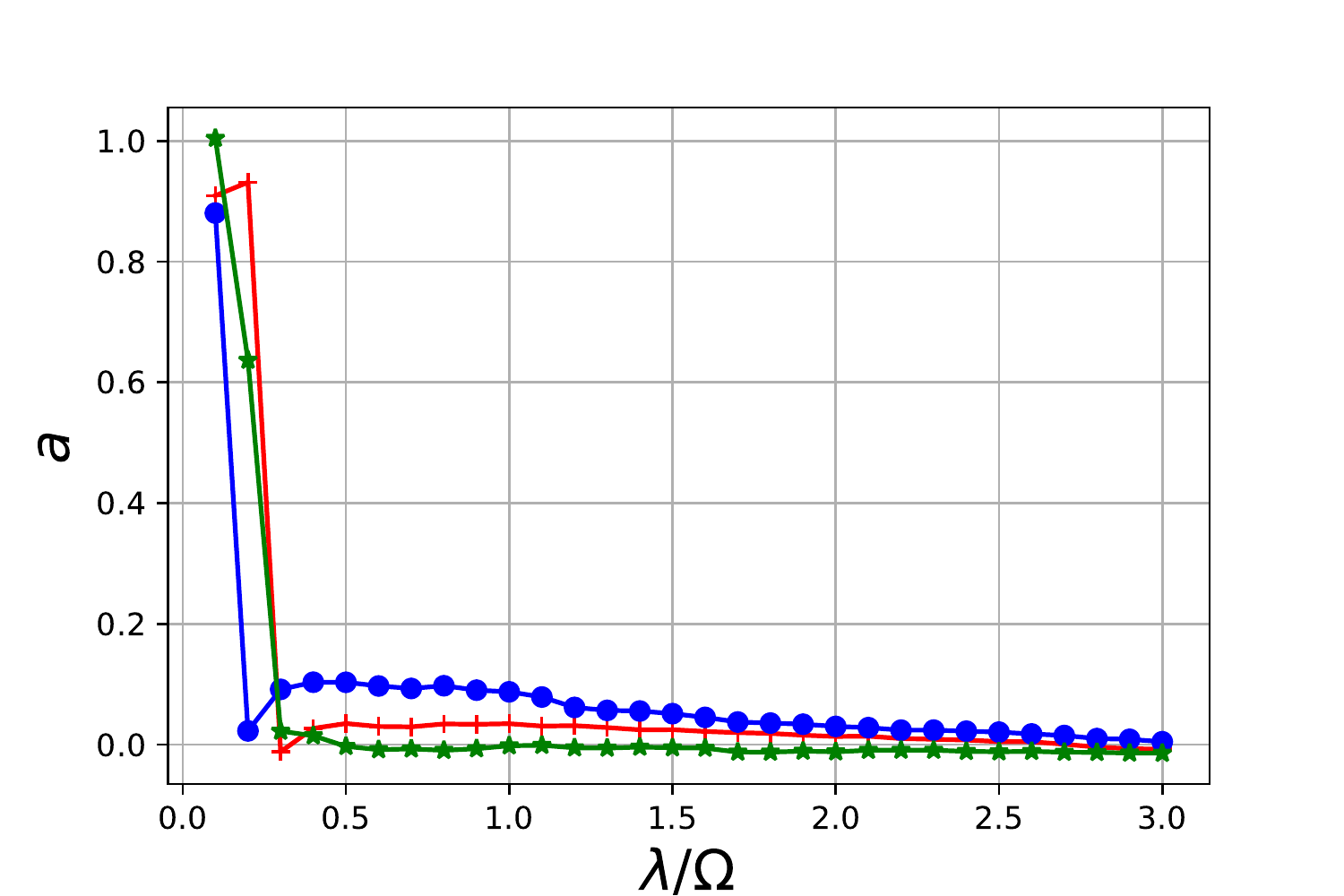}  
\caption{Synchronization diagrams for the Erdos-Renyi network described in Section \ref{subsection}.
The blue circles, red crosses, and green stars correspond, respectively, to the optimal, the random, and
the worst subset $C$ of forced nodes,  see the text for further details. 
Top panels: $r$ and $a$   as functions of
$fF/\Omega$, for $\lambda/\Omega  =2$. The critical external force $F_{\rm c}$, see (\ref{F_c}), 
corresponds precisely to
$fF/\Omega=1$ in this case. As one can see, for $F>F_{\rm c}$, all the oscillators follow the same pace
of the external force since $a \approx 1$. One can  also appreciate that $r$ is indeed
enhanced according to our optimization procedure for $F>F_{\rm c}$. Middle panels:  $r$ and $a$   as functions of
$\lambda/\Omega $, for $fF/\Omega  =2$. Although one sees that synchronization ($r\approx 1$) can occur for
some small values of $\lambda/\Omega $, the global synchronization ($a \approx 1$) does require a larger value for the coupling constant. However, we can see that the threshold values of $\lambda/\Omega $ for the
onset of global synchronization are also compatible with our optimization procedure, in the sense that the smallest threshold corresponds to the optimal set, and the largest to the worst one.  Bottom panels:  $r$ and $a$   as functions of
$\lambda/\Omega $, for $fF/\Omega  = 0.2$. This is a situation where the system is rather intensive to the
external force. The synchronization diagrams are similar for the three cases. Notice, in particular, that 
$a \approx 0$, meaning that the synchronized state does not follow the pace of the external force. This situation is essentialy the same one of the $F=0$ case discussed in \onlinecite{optimal}.
}
\label{Fig2}
\end{figure*}
The figure shows the $(\omega_i,\theta_i(\tau)\mod 2\pi)$ graphics, for a fixed $\tau$
 in a globally
 synchronized regime  with $\lambda/\Omega = 2$ and $fF/\Omega = 3$. Each point in the graphics corresponds
 to a Kuramoto oscillator in the network. The depicted line is the ordinary linear regression of the data.
   The linear correlation between
 $\omega_i$ and $\theta_i$, as incorporated in the ansatz (\ref{Gottwald}), is evident. Moreover, we can clearly 
 identify two displaced oscillator populations with similar linear regression slopes. We will return to this interesting dynamical behavior in the
 last section.

In order to test our optimization scheme, we have considered several synchronization diagrams of the
type $r(\tau)$ and $\psi(\tau)$ versus $\lambda/\Omega$ and    $F/\Omega$, for $\tau$ sufficiently large  to assure the relaxation of any transient regime. The order parameters  $r(\tau)$ and $\psi(\tau)$ are
evaluated accordingly  their definition (\ref{order_parameter}) for the numerical solutions
of (\ref{kuramoto-tau}), taking into account the usual statistical 
precautions, see \onlinecite{PintoSaa,optimal} for further details.  Our results are summarized in
Fig. \ref{Fig2}, from
where one can appreciate that the selection of the subset of forced nodes according to our optimization
scheme  effectively gives origin to networks with better global synchronization capabilities.  The
diagrams $r \times fF/\Omega $ and $r \times \lambda/\Omega $ have
evident meaning and can be easily understood. 
The case for the order parameter $\psi$ is more involved. Its 
asymptotic behavior for large values of $\tau$ in a synchronized regime is expected to be
$
\psi(\tau) \sim a\tau.
$
For a globally synchronized state, we will have $a = 1$, or $\psi\sim \Omega t$, in terms
of the original time variable $t$ of (\ref{kuramoto}). However, different values for $a$ are
also associated with synchronized states, but for which the oscillators do not follow the same
pace of the external force. In particular, the usual synchronization state for the Kuramoto
network with $F=0$ is known to be such that $a = \langle\omega \rangle$. We have explicitly compared three
cases for the subset $C$, namely the optimal, the random, and the worst cases. The optimal and the worst cases correspond, respectively, to selection of the forced nodes as those ones with maximal and minimal 
values of $|\omega_i -\Omega|$. The results are in complete agreement with the expectations for our optimization
scheme: the optimal subset   always implies better global synchronization capabilities than the random one, while the worst subset always exhibits worse capabilities when compared with the random case. 

\section{Final Remarks} 

Although all our analyses have  effectively been  done for the case with $\Omega>0$ and for symmetric distributions
$g(\omega)$ with null average, 
the optimization criterion of selecting the subset $C$ as those oscillators with maximal value of $|\omega_i-\Omega|$ is also valid for the general case. Let us consider, first, the case with $\Omega < 0$. 
From (\ref{kuramoto}), supposing that  $\theta_i(t)$ is a solution for the case with $\Omega>0$ for a network
of Kuramoto oscillators with natural frequencies $\omega_i$, we have that
 $-\theta_i(t)$ will be the solution for
the case corresponding to  $\Omega<0$, but for a network with natural frequencies $-\omega_i$. Both situations will have the same fixed points for the equations equivalent to (\ref{eqalin}) and (\ref{eqblin}), but now the criterion of minimal $\langle \omega \rangle_C$ for $\Omega < 0$ will  
effectively  select those nodes with larger natural frequencies, which indeed corresponds to maximize $|\omega_i-\Omega|$. The case of 
non-symmetric distributions is a bit more involved since one needs to keep track of the terms
proportional to  $\langle \omega \rangle $ in the derivation of (\ref{eqa}) and (\ref{eqb}) and all
the subsequent manipulations. For this case, Eq. (\ref{eqa}) and (\ref{eqb})  will read
\begin{equation}
\label{eqqq}
M \left(\begin{array}{c}
\dot\alpha \\
\dot\beta 
\end{array} \right) = \left(\begin{array}{c}
\langle \omega \rangle - \Omega -  F{\cal I}_1 \\
 \frac{\langle \omega^2 \rangle}{\Omega } - \langle \omega \rangle  - F{\cal I}_2 + \lambda {\cal I}_3  
\end{array} \right) , 
\end{equation} 
where 
\begin{equation}
M = 
\left(\begin{array}{cc}
1 & \frac{\langle \omega \rangle}{\Omega}\\
\frac{\langle \omega \rangle}{\Omega}&  \frac{\langle \omega^2 \rangle}{\Omega^2}
\end{array} \right).
\end{equation}
Since $\det M = \frac{\sigma^2_\omega}{\Omega^2}$, we see that, unless all $\omega_i$ be equal, a situation where our optimization procedure obviously does not apply, the fixed points $(\alpha_*,\beta_*)$ in this case correspond to the zeros of the right-handed side of (\ref{eqqq}). However, the linearization
of the right-handed side of (\ref{eqqq})
around $\beta = 0$ gives origin to   exactly the same condition (\ref{linear}), and hence the same analysis of the symmetric case does apply here. Hence, our optimization criterion is indeed valid for the  case of more
general $\Omega$ and $g(\omega)$.

The role played by possible symmetries of a given   network in its synchronization capability was
recently discussed in \onlinecite{AISync}, for the case of multilayer networks of Stuart-Landau complex
oscillators,  in the context of the so-called  
asymmetry-induced synchronization (AISync) scenario\cite{
NishMotter, ZhangNishMotter, ZhangMotter}.  
The main conclusion  
was   that the 
presence of certain regularities in the interlayer connection
pattern tends to diminish the synchronization capability of the network or, in other words, asymmetries in
the network tend to enhance its synchronization properties. The key point here is  the quadratic quantity 
$\cal L$ given by (\ref{cal_L}), which always decreases if interlayer  symmetries are present, see \onlinecite{AISync} for the details. It is important to stress that the same conclusions hold
here: the presence of possible symmetries as those ones discussed in \onlinecite{AISync} tends to 
diminish the global synchronization capabilities of the forced network. It is also interesting to notice
that, in the autonomous formulation of the problem (\ref{kuramoto}), in which  an extra extra  $i=N+1$ node  with natural frequency $\Omega$ is connected in a directional way to the forced nodes $i\in C$, one
can interpret our optimization scheme in the same way of the optimal synchronization for the $F=0$ case\cite{optimal}, in particular that anti-correlation between the frequencies of neighbor nodes 
always favors synchronization, see also \onlinecite{skardal2014}. The investigation of such phenomena
 might shed some light on the evolution
and structure of natural systems for which global 
synchronization is a desired property.  Many recent studies 
have been devoted for theses issues as, for instance, 
the study of the response to external stimuli
of the C. elegans neural network\cite{modular, p8}, 
self-organization and pattern formation
in the growing and developing of vertebrates\cite{p9},  auditory signals in amphibians\cite{p10}, and several topics on the
circadian rhythm\cite{p11,p12,p13}.  

We finish recalling the dynamical behavior involving the two oscillator populations we could identify in the globally synchronized regime, see Fig. \ref{Fig1}. Typically, the displaced oscillators are the forced ones, giving origin to a curious dynamical configuration. When the global synchronization is completely attained, the forced nodes clump together and move ahead of the remaining swarm of oscillators. The dynamics in this case consist in  a large group of oscillators (the free ones) chasing the smaller group of the forced ones.
The existence of the two disjoint populations prevents the globally synchronized state to have an order parameter $r$ very close to 1, which can be clearly seen in the top left panel in Fig. \ref{Fig2},
where one can observe a sudden decrease in $r$ when global synchronization is attained.  
 It is certainly 
worth to incorporate
such two populations behavior in a second order approximation for the collective ansatz (\ref{Gottwald}). These
points are now under investigation.

\appendix
\section{Mean-field approximations for the random network} 
\label{appendix}
The mean-field expression (\ref{aprox1}) is obtained directly from the evaluation of
(\ref{I_1}). Notice that
\begin{equation}
{\cal I}_1(\alpha,\beta) = \frac{f}{N_c}\sum_{i\in C}  \sin\left( \alpha + \frac{\omega_i}{\Omega}\beta  \right),
\end{equation}
where $f=N_C/N$ is the fraction of forced nodes and the sum is performed on the  
forced nodes. With the hypothesis that $C$ is a random subset of the   oscillators and that
the natural frequencies of the elements in $C$ have the same distribution
$g(\omega)$, one can write
\begin{equation}
{\cal I}_1(\alpha,\beta) =  f\int d\omega\, g(\omega)  \sin\left( \alpha + \frac{\omega}{\Omega}\beta  \right),
\end{equation}
and then (\ref{aprox1}) follows from the definition (\ref{rfixed}) and that $g(\omega)$
is symmetric and has null average. The same hypothesis of a random subset $C$ allows us to write (\ref{I_2})
as
\begin{eqnarray}
{\cal I}_2(\alpha,\beta) &=& f\int d\omega\, g(\omega) \frac{\omega}{\Omega} \sin\left( \alpha + \frac{\omega}{\Omega}\beta  \right) \nonumber \\
&=& f \cos\alpha \int d\omega\, g(\omega)  \frac{\omega}{\Omega} \sin  \frac{\omega}{\Omega}\beta    ,
\end{eqnarray}
where the condition of symmetric $g(\omega)$ with null average was used again, and hence (\ref{aprox2}) follows directly
from (\ref{rfixed}) as well. The evaluation of (\ref{aprox3}) a little bit more involved. First, notice
that
\begin{equation}
\label{I30}
{\cal I}_3( \beta) = \frac{1}{N}\sum_{i=1}^N B_i,
\end{equation}
where
\begin{equation}
\label{I31}
B_i = \frac{\omega_i d_i}{\Omega} \frac{1}{d_i} \sum_{j \in D_i}  \sin \frac{\beta}{\Omega}\left(  \omega_j - \omega_i  \right),
\end{equation}
with $d_i = \sum_j A_{ij}$ standing for the degree of the $i^{\rm th}$-node, which
is a positive integer since our networks are always assumed to be connected.  
The sum in (\ref{I31}) is performed over the set $D_i$ of nodes connected to the $i^{\rm th}$-node 
in the network. Since we are dealing with  random networks with connection pattern independent of the nodes
natural frequencies, one has
\begin{equation}
\label{I32}
B_i = -\frac{\omega_i d_i }{\Omega} \sin \frac{\beta\omega_i}{\Omega} 
\int d\omega\, g(\omega) \cos    \frac{\beta\omega}{\Omega} ,
\end{equation}
where we have used again that $g(\omega)$ is symmetric and has null average. Inserting (\ref{I32}) in (\ref{I30}) leads to
\begin{equation}
{\cal I}_3(  \beta) =  - \langle d  \rangle 
\int d\omega' \,  g(\omega') \frac{ \omega'}{\Omega}\sin    \frac{\beta\omega'}{\Omega} \int d\omega\, g(\omega) \cos    \frac{\beta\omega}{\Omega},
\end{equation}
where $\langle d  \rangle $ is the average degree of the network and we have used the
condition that $d_i$ and $\omega_i$ are independent random variables. Eq. (\ref{aprox3}) follows
now directly from the definition of $r(\beta)$ given by (\ref{rfixed}).

\section*{Acknowledgment}
The authors 
acknowledge the financial support of CNPq    and FAPESP (Grant 2013/09357-9).
They also 
wish to thank M.A.M de Aguiar and C.A. Moreira  for
enlightening discussions and the anonymous referees for valuable suggestions.

\end{document}